# *Neuronal avalanches as a predictive biomarker of BCI performance: towards a tool to guide tailored training program*

Camilla Mannino[1], Pierpaolo Sorrentino[2,3], Mario Chavez[1], Marie-Costance Corsi[1]

[1]Sorbonne Université, Institut du Cerveau – Paris Brain Institute -ICM, CNRS, Inria, Inserm, AP-HP, Hôpital de la Pitié Salpêtrière, F-75013, Paris, France
[2]Institut de Neurosciences des Systèmes, Aix-Marseille Université, 13005 Marseille, France
[3]University of Sassari, Department of Biomedical Sciences, Viale San Pietro, 07100, Sassari, Italy

*Abstract*
Brain-Computer Interfaces (BCIs) based on motor imagery (MI) hold promise for restoring control in individuals with motor impairments. However, up to 30% of users remain unable to effectively use BCIs—a phenomenon termed "BCI inefficiency." This study addresses a major limitation in current BCI training protocols: the use of fixed-length training paradigms that ignore individual learning variability. We propose a novel approach that leverages *neuronal avalanches*—spatiotemporal cascades of brain activity—as biomarkers to characterize and predict user-specific learning mechanism. Using electroencephalography (EEG) data collected across four MI-BCI training sessions in 20 healthy participants, we extracted two features: avalanche length and activations. These features revealed significant training and task-condition effects, particularly in later sessions. Crucially, changes in these features across sessions (Δavalanche length and Δactivations) correlated significantly with BCI performance and enabled prediction of future BCI success via longitudinal Support Vector Regression and Classification models. Predictive accuracy reached up to 91%, with notable improvements after spatial filtering based on selected regions of interest. These findings demonstrate the utility of neuronal avalanche dynamics as robust biomarkers for BCI training, supporting the development of personalized protocols aimed at mitigating BCI illiteracy.

*Introduction*

Motor imagery (MI)-based Brain-Computer Interfaces (BCIs) are promising technologies for restoring communication and control in individuals with severe motor impairments. Despite their potential, a persistent limitation remains: a substantial proportion of users—estimated at 15–30%—fail to gain effective control over the interface, a phenomenon commonly referred to as "BCI inefficiency" (Thompson et al., *2018*[1]). One underappreciated contributor to this challenge lies in the uniform design of BCI training protocols, which typically prescribe a fixed number of sessions for all participants. This "one-size-fits-all" approach overlooks the individual mechanisms underlying BCI performance and neural learning, assuming a homogeneous adaptation curve across users. Considerable inter-individual variability exists in neural plasticity (Dardalat et al, 2023[2]), cognitive strategy, and brain dynamics, which implies that users follow distinct learning trajectories. While some individuals quickly internalize the required mental strategies, others may require longer or more personalized training to induce the necessary neurophysiological adaptations (Alkoby et al., 2018[3]).

A growing body of evidence suggests that both psychological and neurophysiological factors contribute to BCI performance variation, manifesting both across users (inter-subject variability) and within individuals over time (intra-subject variability). BCI performance is modulated by human factors and cognitive traits such as motivation, emotional state, or the strategy use. Higher motivation and positive mood have been associated with enhanced performance (Nijboer et al., 2008, 2010[4,5]), while frustration and overconfidence can impair it (Guger et al., 2003[6]; Witte et al., 2013[7]). Passive or emotionally driven strategies tend to outperform effortful cognitive techniques (Kober et al., 2013[7]; Nan et al., 2012[8]; Hardman et al., 1997[8]), and traits such as visuo-motor coordination, attentional control, imagination ability, and musical training have been linked to better outcomes (Hammer et al., 2012[9]; Vuckovic & Osuagwu, 2013[10]).

On the neurophysiological side, several biomarkers have been proposed to predict BCI aptitude. Resting-state EEG markers—particularly higher alpha power—are consistently associated with better MI control, whereas BCI-illiterate individuals often show increased theta and decreased alpha activity (Blankertz et al., 2010[11]; Ahn et al., 2013b[12]). Functional activation in regions such as the supplementary motor area and parietal cortex also differentiates high from low performers (Halder et al., 2011;[12] Guillot et al., 2008[13]), and structural attributes including white matter integrity and mid-cingulate cortex volume correlate with BCI performance (Halder et al., 2013[14]; Enriquez-Geppert et al., 2014[15]). Event-related desynchronization (ERD), a core signature of motor imagery, tends to be diminished in low-performing users (Sannelli et al., 2019[16]), while dynamic features such as fluctuations in frontal gamma and preparatory alpha power reflect intra-individual variability across sessions (Grosse-Wentrup et al., 2011[17]; Maeder et al., 2012[18]). Importantly, early training performance may itself serve as a predictive marker: Neumann and Birbaumer (2003)[17] showed that early slow cortical potentials (SCP) modulation correlated with later performance in severely paralyzed patients, and Weber et al. (2011)[19] found that sensory motor rhythms (SMR) modulation success in early sessions predicted long-term outcomes, although reliable prediction required at least eleven sessions.

While these findings highlight the utility of early and resting-state EEG features, a major limitation of current BCI paradigms is their reliance on univariate, localized measures—treating brain regions as isolated signal sources. This perspective neglects the fact that brain function emerges from dynamic interactions across distributed neural networks. Accordingly, functional connectivity (FC), defined as the statistical interdependence between brain signals from different regions, has gained attention as a system-level predictor of BCI performance (Mohanty et al., 2020[20]; Gonzalez-Astudillo et al., 2021[21]). However, challenges remain in capturing the temporally evolving nature of these interregional interactions (Mohanty et al., 2020[20]).

Moreover, most BCI studies have historically focused on periodic, oscillatory signals—such as mu and beta rhythms—generated by synchronous neural activity. In contrast, aperiodic signals have often been dismissed as background noise, despite evidence suggesting they encode essential information about the underlying state of neural populations. Recent research indicates that both sustained oscillations and transient burst events coexist in the brain, each providing unique insights into cognitive and motor processes (Brake et al., 2024[22]). A compelling framework for capturing these dynamics involves neuronal avalanches—cascades of burst activity that propagate across cortical networks (Beggs & Plenz, 2003[23]; Arviv et al., 2019[24]). Detected by identifying large signal excursions in coarse-sampled EEG or MEG data, neuronal avalanches offer a spatiotemporal representation of how neural activity initiates and dissipates across brain regions (Sorrentino et al., 2021[25]). These cascades have been shown to preferentially travel along white matter tracts and to shape spontaneous fluctuations in resting-state brain activity (Rabuffo et al., 2021[26]).

Recent work has demonstrated that features derived from Avalanche Transition Matrices (ATMs)—which encode the probability of activity propagation between brain regions—can reliably distinguish between motor imagery and rest conditions in BCI tasks. These features have shown improved inter-subject consistency and better interpretability than conventional oscillatory power metrics (Corsi et al., 2024[27]; Mannino et al., 2024[28]). Unlike traditional features, which reflect the magnitude of localized activity, avalanche-based metrics capture how activity propagates across neural circuits over time—providing a network-level, dynamical account of brain function. This distinction is crucial for understanding individual variability in learning and performance.

Building on the idea that learning induces measurable changes in brain activity, we investigate whether the dynamics of neuronal avalanches can serve as robust biomarkers to capture these changes and guide the development of personalized training programs in MI-based BCI training. We hypothesize that features derived from avalanche propagation—namely avalanche length and spatial activation—encode both task-specific effects and longitudinal changes associated with learning. By capturing the evolution of brain-wide activity patterns, these features may reveal whether a user is adapting effectively and could therefore support early prediction of future performance. Ultimately, this approach seeks to move beyond standardized protocols and toward neurophysiologically informed, individualized BCI systems designed to overcome the problem of BCI illiteracy.

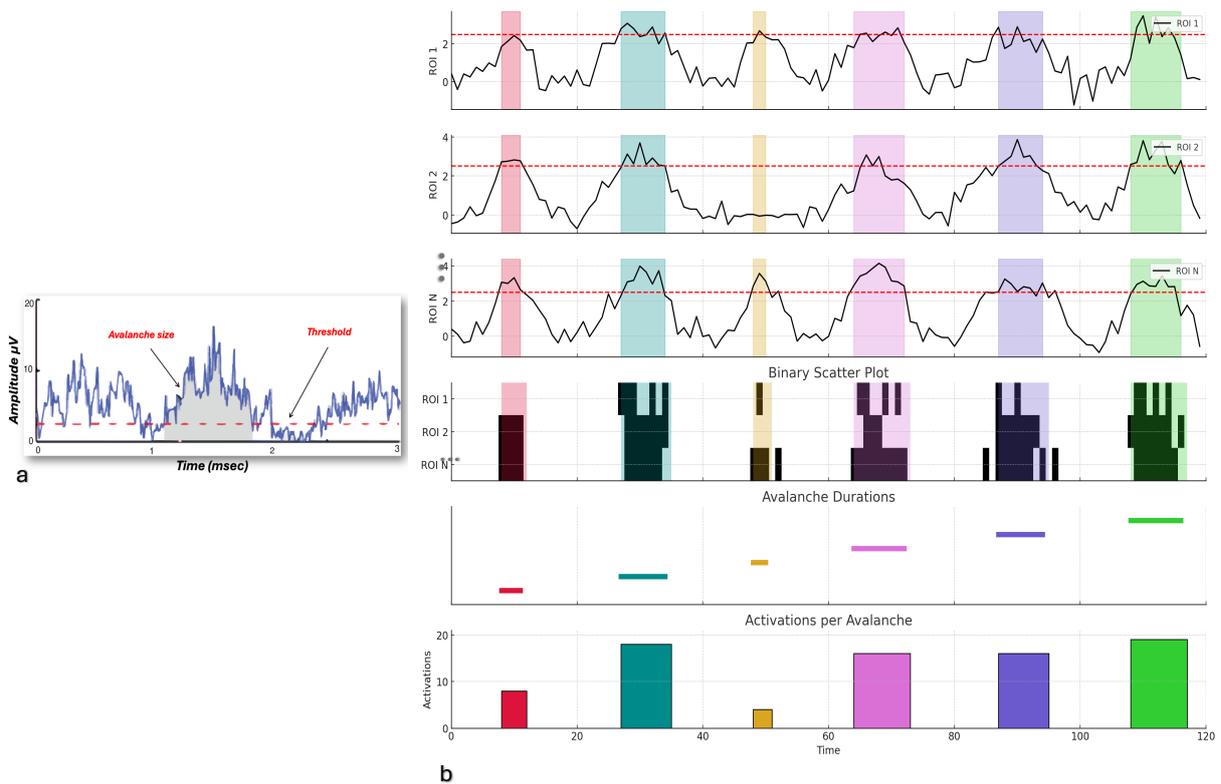

*Figure 1:*
*a)* Schematic representation of a neuronal avalanche: an avalanche begins and ends when the integrated network activity crosses a threshold. It is considered a valid avalanche if it remains above this threshold for at least a predefined minimum duration. *b)* Overview of our method implementation: the first three rows represent data from three different regions of interest (ROIs). Different shades indicate distinct neuronal avalanches, which can span multiple ROIs. The fourth row shows a binary activity scatter plot within each avalanche. The last two rows depict extracted features: the mean avalanche length across avalanches for each trial (fifth row), and the mean number of activations within each avalanche across all ROIs for each trial (sixth row).

*Materials and Methods:*

*2.1   Participants, Experimental Protocol, and EEG Acquisition*

Twenty healthy, right-handed adults (mean age: 27.5 ± 4.0 years; 12 males), all BCI-naive and free of any neurological or psychological disorders, participated in the study. According to the declaration of Helsinki, written informed con- sent was obtained from subjects after explanation of the study, which was approved by the ethical committee CPP-IDF-VI of Paris. All participants received financial compensation for their participation. The experiment followed a longitudinal design, consisting of four EEG-based BCI training sessions over two weeks (two sessions per week). Between sessions, participants were instructed to continue training independently at home using short instructional videos.

The BCI task used in this study was a one-dimensional, two-target cursor control paradigm. To move the cursor toward the upper target, participants performed sustained motor imagery (MI) of right-hand grasping; to reach the lower target, they remained at rest,randomly and equally distributed between upper and lower positions. Each trial began with a 1-second inter-stimulus interval (ISI), followed by a 5-second target presentation. Visual feedback was displayed between 3 and 6 seconds, showing a cursor that started in the center-left of the screen and moved rightward at constant speed. Participants were tasked with controlling the vertical position of the cursor through brain activity modulation.

EEG signals were recorded using a 74-channel Easycap system (Ag/AgCl electrodes) arranged according to the international 10–10 system. Data acquisition took place in a magnetically shielded room with a sampling rate of 1 kHz and a 0.01–300 Hz bandpass filter. Artifact correction was performed using ICA (Infomax), and data were downsampled to 250 Hz. Trials were segmented into 7-second epochs and manually cleaned based on variance and visual inspection, ensuring no more than 10% trial rejection. Source reconstruction was conducted using individual BEM head models and weighted minimum norm estimation (wMNE), with activity mapped to the MNI template and analysed using the Destrieux atlas.
For a complete description of the experimental design, data acquisition, and preprocessing pipeline, please refer to Corsi et al., 2020[29].

*2.2   Features extraction*
*2.2.1   Pre-selection parameters*

EEG remains one of the most widely used modalities for real-time brain-computer interface (BCI) applications due to its non-invasive nature and high temporal resolution. It's why we work whit this data acquisition technique. Since EEG signals are continuous and coarsely sampled, the detection of neuronal avalanches involves discretizing the signal into binary events. This is typically achieved by z-scoring the signal and applying a threshold, with an avalanche defined as a sequence that begins when at least one brain region exceeds this threshold and ends when all regions return to inactivity (Sorrentino et al., 2021[30]).

To enable the detection of neuronal avalanches, it is essential to define two key parameters: the z-threshold and the minimum avalanche duration. The z-threshold determines the signal amplitude above which an event is considered part of a neuronal avalanche. Modifying this value can significantly impact the sensitivity of avalanche detection, thereby influencing the amount and reliability of information extracted from the signal.

The second parameter, the minimum avalanche duration, sets the lower limit on the temporal extent of an avalanche for it to be considered valid. This criterion ensures that only events with sufficient temporal propagation are included in the analysis.

Careful optimization of these parameters is crucial to ensure that the detected neuronal avalanches reflect meaningful neurophysiological activity rather than artifacts. This process helps minimize the influence of non-neural noise sources, such as muscle artifacts or eye-movement-related events, thereby improving the validity of the derived features.

To determine the minimum possible duration of neuronal avalanches, we grounded our selection in established neurophysiological knowledge regarding brain region activation during motor and motor imagery tasks. Specifically, we evaluated three durations—5 ms, 50 ms, and 80 ms—each corresponding to distinct phases of motor-related neural processing.

A duration of 5 ms (approximately 2-time samples at a 250 Hz sampling rate) captures the very early phase of sensory processing, reflecting rapid neuronal responses to sensory input or preparatory signals from structures such as the thalamus or sensory cortex. This brief interval is associated with fast synaptic activation and marks the onset of sensory or motor preparation (Hannah et al., 2018)[31].

A duration of 50 ms (approximately 12-time samples) corresponds to the activation of premotor and supplementary motor areas, which are responsible for formulating the motor plan required for movement initiation. This stage reflects the engagement of neural circuits involved in coordinating muscle activity (Economo et al., 2017[31]).

Finally, a duration of 80 ms (approximately 20-time samples) reflects the activation of the primary motor cortex (M1), at which point the cortex begins transmitting motor commands to the spinal cord. This marks the transition from planning to motor execution (Endo et al., 1999[32]).

For selecting the minimum signal excursion considered relevant—referred to as the z-threshold—we based our parameter choice directly on the signal itself, using the normalized values of the different signals (zero mean and unit variance) as a reference. So, for each subject we evaluate as possible z-thresholds:

$[\mu, \mu+\sigma, \mu+2*\sigma, \mu+3*\sigma, \mu+4*\sigma, \mu+5*\sigma]$

*$\mu$ = mean of the signal on entire dataset;*
*$\sigma$ = standard deviation of the signal on entire dataset*

We have performed the different computations across all subjects, and we have observed a consistency of these values. In a preliminary analysis, we have evaluated in which combinations there are a consistent number of neuronal avalanches and the neuronal physiological validity of their duration (>5ms). Hence, we have been able to reduce our possible combination of threshold and minimum avalanches duration to 10 possibilities (see Table 1):

| z-threshold | Minimum Avalanche Durations |
|---|---|
| $\mu + \sigma$ | 5 ms, 50 ms, 80 ms |
| $\mu + 2\cdot\sigma$ | 5 ms, 50 ms, 80 ms |
| $\mu + 3\cdot\sigma$ | 5 ms, 50 ms |
| μ + 4·σ | 5 ms |
| μ + 5·σ | 5 ms |

*Table 1 ; Resume of possible valid parameters for z-threshold and min. avalanches length*

### 2.2.2 Features extraction

Neuronal avalanches are composed of discrete activations—time samples exceeding a predefined z-threshold—and the distribution of their durations has been shown to be a key indicator of system dynamics. Based on this, we extracted avalanche length and activation count as two primary features to characterize the brain's dynamic behaviour.

Avalanche length captures the temporal extent of neural propagation. To investigate how this parameter evolves over the course of training, we computed the average duration of avalanches within each trial, followed by averaging across trials for each subject.

The second feature, activation count, was used to assess cortical engagement during both motor imagery (MI) and rest conditions. We analysed binarized avalanche data derived from source-reconstructed EEG signals, mapped onto 68 cortical regions of interest (ROIs) defined by the Desikan-Killiany atlas. Each avalanche generated a spatiotemporal pattern of activity encoded as a binary matrix, where values above the z-threshold were marked as 1—these we define as "activations." For each trial, we quantified the total number of activations across all avalanches. To account for variability in avalanche duration, the activation count for each avalanche was weighted by its temporal length. The overall weighted activation score for a given trial was calculated by dividing the total number of weighted activations (across all time samples and ROIs) by the cumulative duration (in time samples) of all avalanches within that trial.

$$\text{Weigthed mean activations} = \frac{\sum_{av}(\text{Activations count}_{av} \times \text{length}_{av})}{\sum_{av} \text{length}_{av}} \quad (1)$$

*where av indexes the avalanche segments within the trial.*

Once computed, these trial-level activation scores were averaged across all trials for each subject and cardinal position, resulting in a single mean activation score per subject.

We extracted these features for each subject, each condition (Rest and MI) and each of the four sessions.

### 2.3 Statistical analysis and repeated-measures correlation

#### 2.3.1 Statistical analysis

To assess the effects of state (motor imagery vs. rest) and training session (sessions S1–S2-S3-S4) on the two extracted features values across the different possible pre-selected parameters couple, we employed a combination of non-parametric statistical approaches, each tailored to different inferential levels: global versus local effect evaluation.

**Permutation-Based Repeated Measures ANOVA (Global Effects)**
To evaluate the global effects of state, session training effect, and their interaction without relying on normality assumptions, we used a permutation-based approach with 10,000 iterations. In each iteration, the dependent variable was randomly shuffled across subjects and conditions, and F-values were recalculated. Empirical p-values were derived by comparing the observed F-statistics to the distribution of permuted values. This approach allows a global assessment of whether there are systematic effects across all sessions and conditions, rather than focusing on individual pairwise comparisons.

**Local Effects**

To complement the global analysis, we applied non-parametric tests designed for more localized inference:
- Learning effect: To capture within-condition temporal effects, we conduce Friedman Test independently for the MI and Rest conditions to evaluate whether there were significant changes across the four sessions.
- Task condition effect: This test is sensitive to condition-specific differences at specific time points, providing a localized effect perspective. Wilcoxon Signed-Rank Test is independently performed for each session to directly compare the MI and Rest conditions at a per-session level.

### 2.3.2 Repeated correlation

We implement a repeated measure correlation analysis to identify which parameters get a significant correlation with BCI-score and use them as possible candidate for our predict model.

$$\Delta \text{avalanche\_length} = \text{Avalanche\_length}_{Rest(i,S)} - \text{Avalanche\_length}_{MI(i,S)} \quad (2)$$

$$\Delta \text{activations} = \text{Activations}_{Rest(i,S)} - \text{Activations}_{MI(i,S)} \quad (3)$$

*where i indexes subjects and S indexes session.*

This was repeated for each of the preselected couple-parameters.

To estimate the correlations between BCI scores and, respectively, $\Delta$avalanche_length and $\Delta$activations, we performed repeated-measures correlations (Bakdash and Marusich, 2017[33]) which control for non-independence of observations obtained within each subject without averaging or aggregating data. It allows us to get a more accurate estimate of the strength of correlation in longitudinal data than standard Pearson correlation.

### 2.4 ROIs selection

We decide to perform a ROIs selection to evaluate if thanks to it we get a dataset we get a less noise dataset that will show higher correlation and prediction accuracy.

Following the computation of activation values across trials, we derived ROI-level features to analyse spatial patterns of cortical engagement during the BCI training. Each trial was represented as a matrix of binarized source activity over 68 anatomical ROIs, defined according to the Desikan-Killiany atlas. For each trial, the activity within each ROI was aggregated and weighted by segment duration, resulting in a length-weighted mean activation per ROI.

To reduce inter-subject variability and enable comparisons across sessions and conditions, ROI activations were normalized per subject. Specifically, for each subject and ROI, activations were scaled as a percentage of the maximum activation observed during the first Rest session:

$$\textit{Normalized activations } ROI_{i,j} = \left( \frac{ROI\ Activations_{i,j}}{max_j (ROI\ Activation_i)_{Rest\ 1Session}} \right) \times 100 \quad (4)$$

*i is the subject index, j is ROIs index*

Importantly, while we used the maximum as the reference in this study, the normalization could equivalently be based on the mean, median, or minimum ROI activation without substantially altering the outcome. Since all measures were applied within-subject and per ROI, the resulting normalized values preserved relative spatial patterns and trends across conditions, regardless of the specific scaling baseline.

After computing activation values at the single-ROI level for each subject, session, and condition (MI and Rest), we performed within-subject paired t-tests to compare the two states independently for each session. This step allowed us to identify ROIs showing the greatest differences in activation between MI and Rest conditions. The resulting t-value maps—one per subject and session—were then used as input for a two-way ANOVA, to jointly consider task condition effect and learning effect across subjects. This process ultimately yielded a set of significant ROIs that were robust across both individuals and training sessions.

*2.5    Predictive model*

To investigate to which extent the extracted features could be relevant to predict the BCI score in the subsequent session we implemented two models: a regression model to estimate the exact BCI score and a classification model to determine whether the score indicates effective BCI control. For classification, we set the threshold at 57%, which represents the chance level in our dataset (Muller-Putz et al, 2008 [34]); scores above this threshold are interpreted as successful BCI control, while scores below suggest the subject is unable to control the device. Both models take the same inputs—Δavalanche length and Δactivations—as used in the repeated correlation analysis and on both a grid search for hyperparameters is done.

To evaluate model performance, we applied leave-one-out cross-validation (LOO), a strict validation approach particularly suited for small datasets. For each iteration, data from one subject were held out as the test set, while the remaining subjects formed the training set. This process was repeated until each subject had served once as the test case, ensuring that the model was evaluated across all available subjects.

To assess the effectiveness of our models, we compared their performance against two baselines:
1. A standard Support Vector Regression (SVR) (Awad et al., 2015[35]) or Support Vector Classifier (SVC) (Mammone et al.2009 [36]) without temporal modeling.
2. Additionally, using the longitudinal model, we conducted a random shuffling of session order to isolate and quantify the contribution of temporal learning effects (called random sessions).

**Longitudinal Support Vector Regression (LSVR)**

We implemented a Longitudinal Support Vector Regression model, inspired by the framework of Du et al. (2015)[37]. In this approach, we assume repeated measurements for each subject, i, over multiple time points, represented as a list of subject-specific matrices $X_i \in \mathbb{R}^{S \times f}$ and outcome vectors $y_i \in \mathbb{R}^S$, where S is the number of sessions and f is the number of features.

To model temporal trends, we introduce a temporal weight vector $\beta \in \mathbb{R}^S$ that projects both predictors and outcomes into a shared subspace representing longitudinal progression. The regression problem is formulated in the dual space using a quadratic programming (QP) framework, where the objective includes an ε-insensitive loss and dual variables γ∗,γ associated with each subject. The Gram matrix G encodes temporal covariance using $G_{ij} = \beta^T X_i X_j^T \beta$.                                                              (5)

After solving the QP to obtain the dual coefficients α=[γ∗;γ], we estimate the optimal temporal trend vector β by solving a linear system derived from the dual variables and the outcome projections. This iterative estimation allows joint modelling of within-subject dependencies and between-subject variation.

Prediction for new subjects is performed by computing a weighted combination of inner products between the new subject's data and those of the support vectors, modulated by the learned β and dual coefficients.

**Longitudinal Support Vector Classification (LSVC)**
We implemented a Longitudinal Support Vector Classifier (LSVC), extending the standard SVC to account for repeated measurements across sessions, as proposed by Chen and Bowman (2011)[38]. The aim is to jointly estimates the separating hyperplane parameters and the temporal trend parameters using quadratic programming, so a key challenge that we address is how to jointly estimate the parameter vectors β and α. In this framework, longitudinal data are collected from N subjects over S sessions, with features f measured at each session point. For each subject, the data form a matrix $X_i \in \mathbb{R}^{S \times f}$, and the classification label is ys∈{0,1}.
To incorporate the temporal structure, we model each subject's data as a weighted combination of measurements across time:
$$\tilde{x}_i = x_{i,1} + \beta_1 x_{i,2} + \cdots \beta_{S-1} x_{i,S} \quad (6)$$
where $\beta = (1, \beta_1, \beta_2 \ldots \beta_{S-1})$ encodes the temporal trend.
The LSVC algorithm alternates between:
1. Solving the dual SVM problem using quadratic programming with the current β, where the kernel matrix $G_{ij} = \beta^T X_i X_j^T \beta$ incorporates temporal dependencies.
2. Updating β by solving a regularized linear system derived from the current dual coefficients and subject data.

## *Results*

### *3.1 Changes of avalanche lengths and activations across time*

At a global level, as assessed using a permutation ANOVA (red line in Figure 2), no significant effects were observed for either condition (Rest vs. Motor Imagery [MI]) or training (across sessions) in any combination of the coupled parameters (z-threshold and minimum avalanche length). This finding holds for both extracted features: avalanche length and neural activations.

However, a more granular analysis revealed significant within-session differences between the Rest and MI conditions, specifically during the fourth session (blue solid line in Figure 2a and in Figure 2b), for both features using many possible coupled parameters.

In addition, using same coupled parameters, analysis revealed a significant learning effect across sessions within the MI condition (blue solid line in Figure 2c and in Figure 2d) for both features:

All results presented in this study were computed using the broad frequency band [8–35 Hz]. Although not shown here, we also tested additional frequency bands, including the alpha band [8–12 Hz], beta band [12–30 Hz], low gamma band [30–50 Hz], and theta band [3–8 Hz]. In both the alpha and beta bands, the results retained the key characteristics observed in the broad band, although the broad frequency range provided a more comprehensive and integrative view across the various analyses performed. This observation aligns with existing knowledge in the field. Our features directly capture neuronal avalanche dynamics, and prior studies suggest that arrhythmic neural activity can give rise to broadband EEG signals. These findings imply that narrowband EEG power may not always correspond to true oscillatory brain rhythms, reinforcing the relevance of broadband analysis in our context.

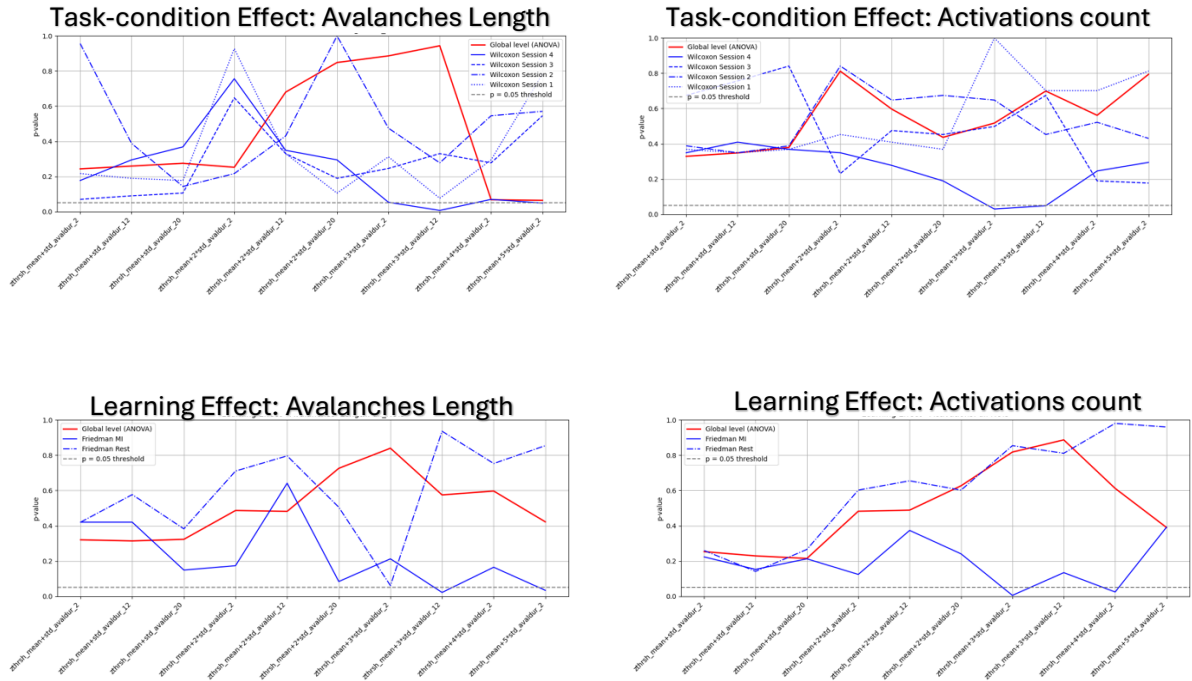

*Figure 2: a) Avalanches Length, Task-Condition Effect: Significance z-threshold $\mu+3\sigma$, min. avalanche length 12: W = 34.00, p = 0.0064 (Wilcoxon Test: MI vs. Rest, Session 4); z-threshold $\mu+5\sigma$, min. avalanche length 2: W = 52.00, p = 0.0484 (Wilcoxon Test: MI vs. Rest, Session 4); b) Activations count, Task-Condition Effect: Significance: z-threshold $\mu+3\sigma$, min. avalanche length 2: W = 47.00, p = 0.0296 (Wilcoxon Test: MI vs. Rest, Session 4); z-threshold $\mu+3\sigma$, min. avalanche length 12: W = 52.00, p = 0.0484 (Wilcoxon Test: MI vs. Rest, Session 4); c) Avalanches Length, Learning Effect: Significance: z-threshold $\mu+3\sigma$, min. avalanche length 12: $\chi^2$ = 9.66, p = 0.0217 (Friedman Test inside MI sessions); z-threshold $\mu+5\sigma$, min. avalanche length 2: $\chi^2$ = 8.70, p = 0.0336 (Friedman Test inside MI sessions); d) Activations count, Learning Effect, Significance: z-threshold $\mu+3\sigma$, min. avalanche length 2: $\chi^2$ = 12.78, p = 0.0051 (Friedman Test inside MI sessions)*

## 3.2 Repetead Correlation

Using the same combinations of coupled parameters, we also observed significant correlations between the changes in avalanche length (Δavalanche length, Figure 3a) and neural activations (Δactivations, Figure 3b) with the BCI performance scores. Specifically, significant positive correlations ($p < 0.05$) were found in different conditions (see Figure 3).

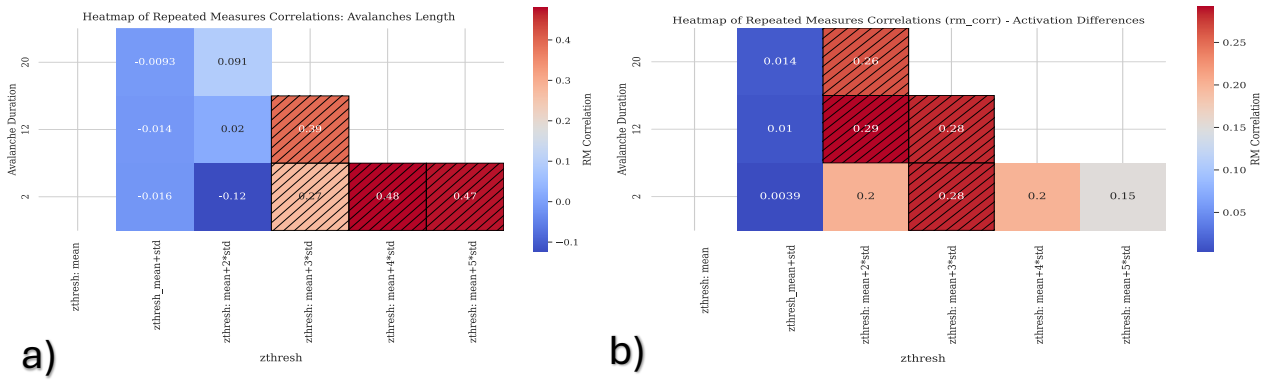

*Figure 3: a) Repeated correlation Δavalanches length [Rest - MI] -BCI-score : z-threshold: $\mu + 3\sigma$, min. avalanche length: 2 — r = 0.27, p = 0.03317; z-threshold: $\mu + 3\sigma$, min. avalanche length: 12 — r = 0.39, p = 0.00181 z-threshold: $\mu + 4\sigma$, min. avalanche length: 2 — r = 0.48, p = 0.000089; z-threshold: $\mu + 5\sigma$, min. avalanche length: 2 — r = 0.47, p = 0.00019; b) Repeated correlation Δactivations [Rest - MI] – BCI-score: z-threshold: $\mu + 2$, min. avalanche length: 12 — r = 0.29, p = 0.022; z-threshold: $\mu + 2\sigma$, min. avalanche length: 20 — r = 0.26, p = 0.040; z-threshold: $\mu + 3\sigma$, min. avalanche length: 2 — r = 0.28, p = 0.027; z-threshold: $\mu + 3\sigma$, min. avalanche length: 12 — r = 0.28, p = 0.028; The significant correlations are highlighted using different textures.*

In all cases, the correlations were positive, indicating that as the difference between Rest and MI conditions increased across sessions, the BCI performance also improved. Notably, this progressive difference was associated with a monotonic increase in the strength of the correlation between both Δavalanche length and Δactivations with BCI scores, observed in most participants (see Figure 4a and Figure 4b).

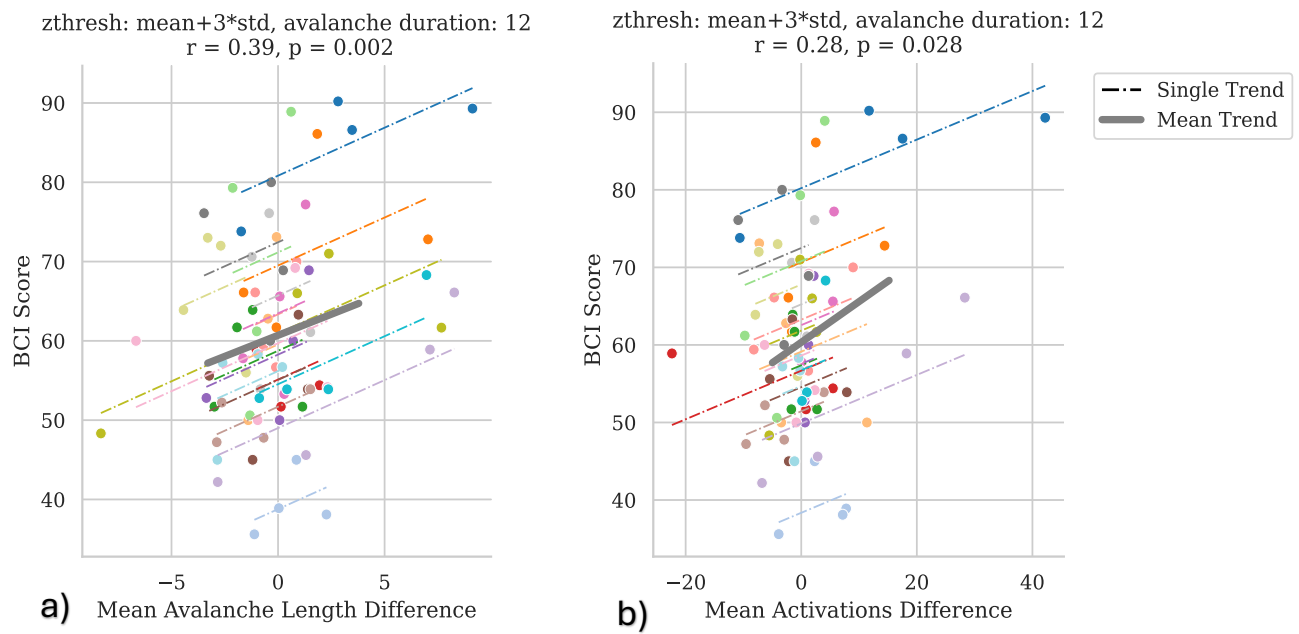

*Figure 4: Repetead correlation trend across different sessions ΔAvalanches length – BCI-score (a); Repetead correlation trend across different sessions ΔAvalanches length – BCI-score (b) each coloured dashed line correspond to one subject while the green bold line identify the trend across all the subjects. We show here the coupled parameters that achieve the best prediction performance.*

### 3.3 Predictive Model

Using the coupled parameters that yielded significant correlations with BCI performance, we implemented two longitudinal predictive models: a regression model (Longitudinal Support Vector Regression, **LSVR**) and a classification model (Longitudinal Support Vector Classifier, **LSVC**). These models were designed to predict each subject's BCI score one session ahead based on data acquired up to the previous session.

Among the possible coupled parameters that show a significant correlation between Δ of our features and BCI-score, the highest predictive performance is achieved thanks to z-thresh equals to $\mu + 3\sigma$ and min. avalanches length equals to 12 (see Figure 4a and Figure 4b).

*Longitudinal Support Vector Regression (LSVR)*

The LSVR model outputs a continuous predicted BCI score, which we evaluated using the Root Mean Square Error (RMSE) against the actual continuous BCI scores. The general trend of the prediction using this model is not optimal even the RMSE result smaller than the RMSE computing using the SVR (RMSE LSVR: 12.7481 vs RMSE SVR: 17.2186), less evident difference is observable comparing LSVR with the shuffling of sessions (RMSE LSVR: 12.7481 vs RMSE LSVR random session: 13.3571).

*Longitudinal Support Vector Classifier (LSVC)*

The RMSE is not enough to allow us to have a measure able to sum-up how close we are to our main goal. Our primary objective was to determine whether a subject would be able to control the BCI system in the following session—based on individual learning progress—or whether additional training would be necessary. For this reason, we also implemented the LSVC model to perform binary classification. We defined successful control using a performance threshold set at the chance level of 57%, as reported by Müller-Putz et al. 2008[34]

Thanks to LSVC we achieve 88% accuracy in predicted performance. Four subjects on twenty total as missed classified (see confusion matrix in Supplementary Materials Figure 1). LSVC model significantly outperformed the standard SVC, achieving 88% compared to 41% (see confusion matrix in Supplementary Materials Figure 1).

Additionally, to evaluate the influence of the learning effect over time, we performed a session-shuffling control analysis. When the temporal order of the three sessions was randomly shuffled, classification accuracy dropped to 47% (see confusion matrix in Supplementary Materials Figure 1).

These results suggest that the longitudinal structure of the data contributes meaningfully to classification performance, although the limited number of sessions (three) may constrain the observable effect.

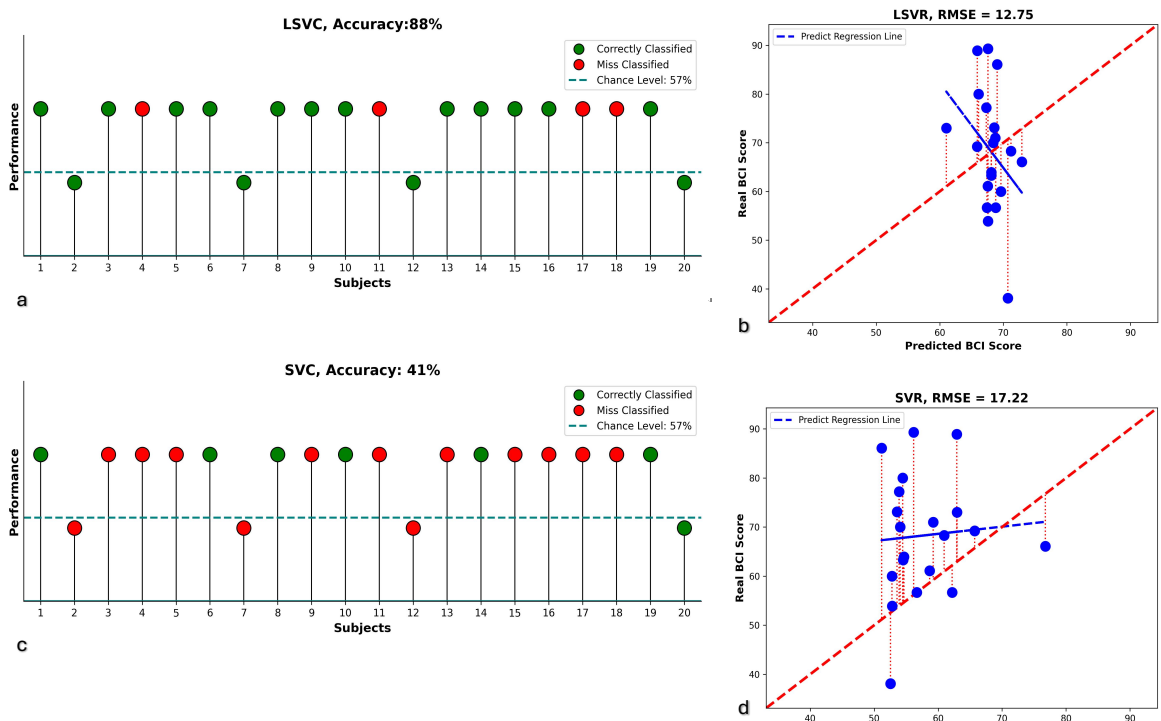

*Figure 5: Predictive model results.*
*(a, c) Classification results*: *Each ball represents one subject. The dashed line indicates the classification threshold; subjects above the line are considered able to control the outcome. Green balls indicate correct predictions of control ability, while red balls represent misclassifications. Panel (a) shows results using a longitudinal predictive model, which are compared to a standard SVC model in panel (c).*
*(b, d) Regression results*: *Each point represents one subject. The red dashed lines indicate the prediction error between actual and predicted values. The bold red line shows the optimal prediction trend, while blue lines represent the predicted trends. Panel (b) shows results from a longitudinal regression model, compared to a standard SVR model in panel (d).*

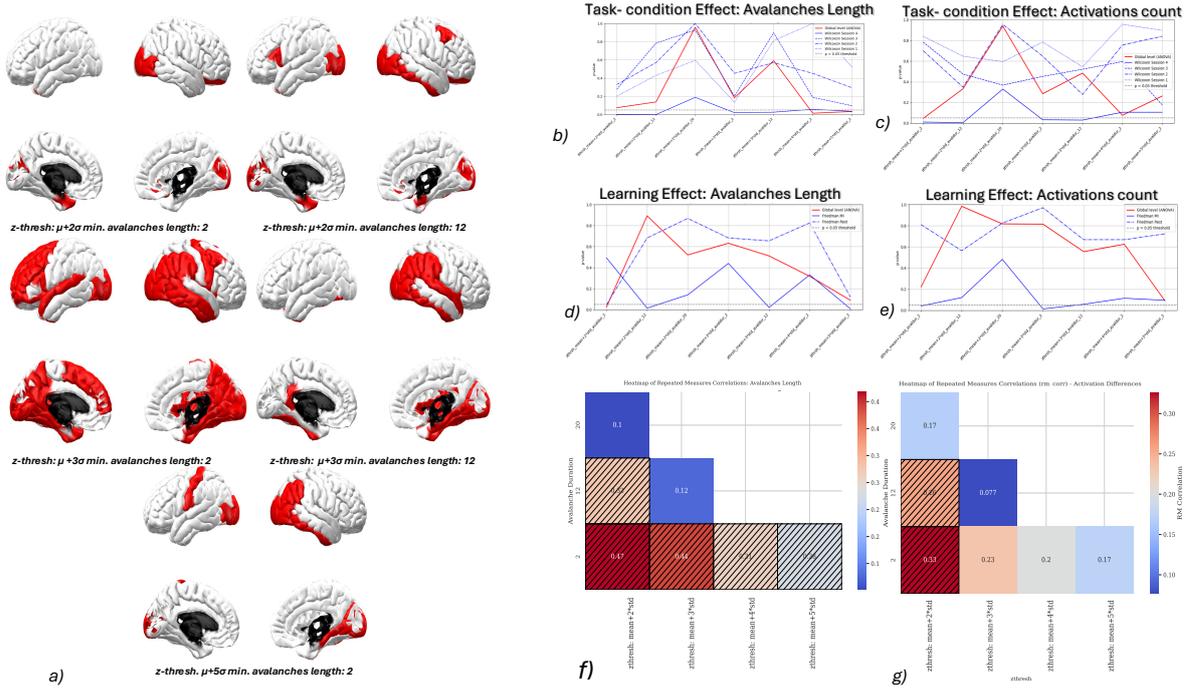

*Figure 6:* a) Selected ROIs set for each coupled parameters, the ROIs with p-value < 0.05 [two-way ANOVA] are in red; b) *Avalanches Length, Task-Condition Effect: Global Level Significance (Rest vs MI across all session):* z-thresh = μ + 5σ, min. avalanche length = 2 — F = 4.4123, $p = 0.0342$; c) *Activations Task-Condition Effect: Global Level Significance (Rest vs MI across all session):* z-thresh = μ + 2σ, min. avalanche length = 2 — F = 4.0135, $p = 0.0453$; d) *Avalanches Length Learning Effect: Global Level Significance:* z-thresh = μ + 2σ, min. avalanche length = 2 — F = 3.1890, $p = 0.0244$ e) *Activations counts Learning Effect: No global level significance;* f) *Significant repeated correlation between Δavalanche length [Rest − MI] and BCI scores over a selected set of ROIs:* z-threshold: μ + 2σ, minimum avalanche length = 2 — r = 0.47, $p = 0.0001$; z-threshold: μ + 2σ, minimum avalanche length = 12 — r = 0.33, $p = 0.0104$; z-threshold: μ + 3σ, minimum avalanche length = 2 — r = 0.44, $p = 0.00041$; z-threshold: μ + 4σ, minimum avalanche length = 2 — r = 0.31, $p = 0.0128$; z-threshold: μ + 5σ, minimum avalanche length = 2 — r = 0.28, $p = 0.0316$; g) *Significant repeated correlation between Δactivations [Rest − MI] and BCI scores over a selected set of ROIs:* z-threshold: μ + 2σ, minimum avalanche length = 2 — r = 0.33, $p = 0.010$; z-threshold: μ + 2σ, minimum avalanche length = 12 — r = 0.26, $p = 0.046$. Significant correlations are visually emphasized in the figure using different textures.

### 3.4 Selected ROIs

Following the ROI selection process described in the *Materials and Methods*, we identified a common set of significant ROIs across all subjects for each parameter combination. This set (see Figure 6a) encodes both the condition-related differences (Rest vs. MI), as reflected in the *t*-values, and the learning effects over time, as captured by ANOVA across sessions.

It is worth noting that the parameter combination using a z-threshold equal to *μ + 1σ* did not yield any significant ROIs. Therefore, in this section, we focus exclusively on parameter combinations with a z-threshold of at least *μ + 2σ*. For each selected set of significant ROIs, we recomputed the neuronal avalanches and extracted the same two key features: avalanche length (see Figures 6b, 6d, and 6g) and activations (see Figures 6c, 6e, and 6g).

Using only these selected ROIs, we were able to detect both local and global statistical effects. Locally, we observed significant differences between Rest and MI conditions during the fourth session, particularly for avalanche length and activation features (Wilcoxon test, *p* < 0.05; see blue solid lines in Figures 6b and 6c). Additionally, a significant learning effect was present within the MI condition across sessions (Friedman test, *p* < 0.05; blue solid lines in Figures 5d and 6e). Globally, using a 10,000-permutation ANOVA, we also observed significant condition and learning effects (see red lines in Figures 6b–6e).

The repeated correlation analysis between BCI performance and both features mirrored earlier findings and revealed significant trends, see Figure 6.

As in the previous section, we trained two longitudinal predictive models—LSVR (regression) and LSVC (classification)—using only the parameter combinations that showed significant repeated correlations. Best results are obtained using the parameter setting of *z-thresh = μ + 2σ* and *min. avalanche length = 12* for both Δavalanche length and Δactivations (see Figure 7).

**LSVR** **Analysis:**
The LSVR model yielded a slightly lower RMSE (10.3931) compared to the model trained on the full dataset without ROI selection (RMSE = 12.7481) and as can be observed from the plot also the general regression trend fit better with the real trend after the ROIs selection (see Figure 5d). To assess the benefit of incorporating longitudinal information, we compared the LSVR model with the standard SVR model that does not account for session order. LSVR achieved a significantly lower RMSE (10.3931 vs. 17.0468) (Figure 8b, Figure 8d), highlighting the added value of modelling session progression. While a random shuffle of session order perform a slightly difference in RMSE value (10.3931 vs. 13.3219), the results still suggest a meaningful contribution of longitudinal structure.

**LSVC** **Analysis:**
The LSVC model achieved an accuracy of 91%, (see Figure 8a) with only 3 false negatives out of 20 subjects (see Figure 8 and Supplementary Material 2). When compared to the standard SVC model that ignores session order, LSVC outperformed it significantly—achieving 91% accuracy (Figure 8a, Figure 8c) versus 59%. To further investigate the importance of session progression, we conducted a random shuffle of session order. This manipulation led to a drop in LSVC classification accuracy to 84%, suggesting that longitudinal structure plays a notable role. However, given the limited number of sessions (only three), the true extent of this effect may be underestimated.

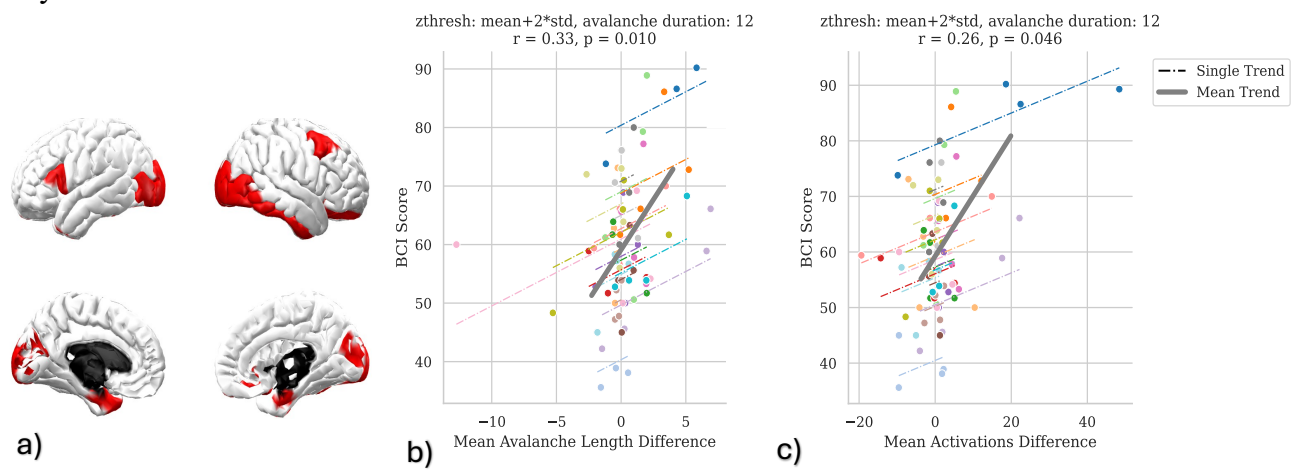

*Figure 7: (a) Selected set of ROIs; (b) Repetead correlation trend over a selected set of ROIs across different sessions ΔAvalanches length – BCI-score; (c) Repetead correlation trend over a selected set of ROIs across different sessions ΔAvalanches length – BCI-score; each coloured dashed line correspond to one subject while the green bold line identify the trend across all the subjects. We show here the coupled parameters that achieve the best prediction performance over a selected set of ROIs.*

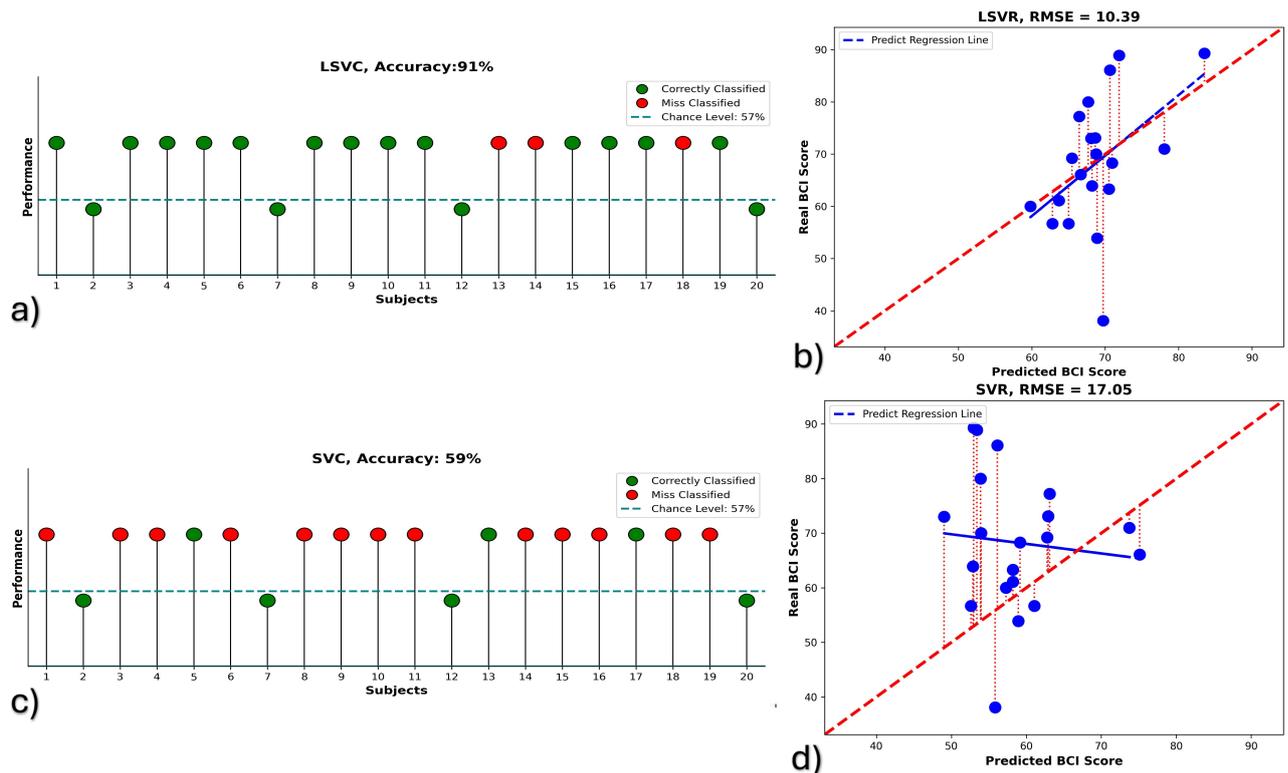

*Figure 5: Predictive model results on a selected set of ROIs*
**(a, c) Classification results:** *Each ball represents one subject. The dashed line indicates the classification threshold; subjects above the line are considered able to control the outcome. Green balls indicate correct predictions of control ability, while red balls represent misclassifications. Panel (a) shows results using a longitudinal predictive model, which are compared to a standard SVC model in panel (c).*
**(b, d) Regression results**: *Each point represents one subject. The red dashed lines indicate the prediction error between actual and predicted values. The bold red line shows the optimal prediction trend, while blue lines represent the predicted trends. Panel (b) shows results from a longitudinal regression model, compared to a standard SVR model in panel (d).*

## Discussion

This study aimed to investigate the neural mechanisms underlying motor imagery-based brain–computer interface (MI-BCI) training, with a particular focus on identifying condition-induced neural modulations and predictors of individual performance. BCI and neurofeedback outcomes are shaped by a complex interplay of both stable and dynamic factors, including psychological traits, cognitive abilities, and neurophysiological patterns.

Performance in BCI tasks is influenced by psychological characteristics, cognitive skills, and individual differences. People with greater confidence in technology use, stronger intrinsic motivation, and positive emotional states generally achieve better results; however, excessive confidence or emotional frustration can negatively impact performance (Ahn et al., 2015[39]; Alkoby et al., 2018[3]). Furthermore, intuitive mental strategies—such as spontaneous engagement or positive emotional imagery—tend to elicit more favorable neural responses. Cognitive abilities like visuo-motor coordination and sustained attention also support effective self-regulation. Personal factors such as gender, musical training, and vivid mental imagery have similarly been associated with enhanced BCI performance. From a neurophysiological perspective, there is evidence of both learning effects and performance predictability based on localized features such as event-related synchronization/potentials (ERS/ERP) and sensorimotor rhythms (SMR). For example, higher resting-state alpha power is associated with better performance, while a high theta-to-low alpha ratio is often observed in lower-performing individuals (Blankertz et al., 2010[11]). Additionally, fluctuations in gamma and alpha band activity have been linked to moment-to-moment variations in attentional engagement, further influencing task success (Ahn et al., 2013[40]).

Complementary findings from fMRI studies support these observations, showing that successful users exhibit stronger activation in motor-related brain regions and structural features such as increased white matter connectivity (Halder et al., 2011[12]). These physiological markers suggest promising avenues for adaptive and personalized training strategies. For instance, Corsi et al. (2020)[29] reported that high-performing users progressively showed stronger sensorimotor activation and decreased connectivity in associative cortical areas—indicating a transition toward more automatic and efficient BCI control. Similarly, Stiso et al. (2020)[41] demonstrated that task-related changes in functional connectivity, particularly within frontal networks, were associated with improved attentional modulation and performance.

Collectively, these findings highlight that successful BCI learning depends not only on localized brain activations but also on a broader reorganization and downregulation of cognitively demanding network interactions. Building on this knowledge, the present study introduces a functional connectivity approach designed to also capture the behaviour of aperiodic neural signals.

## 4.1 Task Condition Effect and Learning Effect

Specifically, we examined two extracted features—avalanche length and activation—and found that the effect of task condition (Rest vs. Motor Imagery, MI) became most pronounced during the fourth session, as indicated by statistically significant differences (Wilcoxon test, $p < 0.05$). Additionally, a more robust learning effect emerged across sessions within the MI condition (Friedman test, $p < 0.05$). The convergence of these findings points to an underlying learning mechanism, wherein the brain gradually adapts to enhance focus on the imagined right-hand grasping task. This adaptation strengthens the distinction between resting and motor imagery brain states, offering further evidence of functional plasticity in MI-BCI training.

This hypothesis is further supported by the analysis of probability density functions (PDFs) of our features across subjects, particularly for avalanche length. We observed a progressive and condition-specific change in neural dynamics throughout the BCI training sessions. Specifically, during the Motor Imagery (MI) condition, there was a consistent increase in the average avalanche length across sessions, as shown by the rightward shift and sharpening of the red probability density functions (PDFs) in Figure 9. In contrast, the Rest condition displayed no significant change, with its blue PDFs remaining largely stable. Within each session, the shape and separation of the PDFs reveal a dynamic learning process: in the first session, Rest and MI distributions nearly overlapped, indicating minimal distinction between brain states. However, with continued training, MI-related neural activity diverged progressively from Rest. By session four, the Rest distribution became broader and flatter—signaling higher inter-subject variability—while the MI distribution grew narrower and more peaked, reflecting a more uniform and consistent neural response among most subjects. Notably, the MI distribution developed a bimodal pattern, indicating the emergence of two distinct participant subgroups. The primary group exhibited significantly longer avalanche lengths than the Rest condition, consistent with improved motor imagery proficiency and neural adaptation. The secondary group, whose MI-related activity remained similar to Rest, included subjects such as 2 and 20, who failed to reach above-chance BCI performance. This divergence underscores the presence of individual differences in the capacity to develop effective MI strategies. These findings align with the known learning effects in MI-based BCI training, where neural activity becomes increasingly structured and distinguishable from rest with practice and task familiarization (Jeunet et al., 2016[42]; Vidaurre et al., 2010[43]).

Our results suggest that applying ROI selection is not only feasible but also beneficial, as it does not compromise the statistical significance of our findings. On the contrary, this approach enhances our ability to detect significant differences between Motor Imagery (MI) and Rest conditions across all sessions, as well as robust learning effects in both brain states. This represents an important foundation for our analysis. We acknowledge that the ROI selection method used in this study is not the most precise, as it generates a distinct set of ROIs for each parameter combination and occasionally includes regions outside the primary motor areas targeted by the BCI—such as occipital regions. However, it is important to note that these non-motor areas, including associative and visual processing regions, could plausibly be involved in the training process, contributing to motor imagery through integrative and supportive cognitive mechanisms. This interpretation is further supported by the experimental design, in which subjects receive three seconds of visual feedback during each trial, likely engaging visual and associative cortices alongside motor areas (Corsi et al. 2020)[29]. Nevertheless, these results highlight a key insight: not all ROIs contribute meaningful information, and some merely introduce background noise. By excluding non-informative ROIs, we enhanced the overall signal-to-noise ratio, enabling us to uncover consistent, global differences between Rest and MI states across sessions, as well as clear learning-related effects.

To validate the effectiveness of our extracted features—avalanche length and activation—we further analysed data from the final session by categorizing each trial for each subject as either a Hit (successful BCI control) or a Miss (unsuccessful BCI control). For several parameter combinations, both features reliably distinguished between MI Hit and MI Miss trials, Rest Hit and Rest Miss trials, as well as MI Hit versus Rest Hit trials, with statistical significance ($p < 0.05$ and $p < 0.005$). Moreover, for most parameter settings that showed strong and consistent correlations with BCI performance (as discussed in the previous section), we observed significant interaction effects (permutation ANOVA, $p < 0.05$). In the case of avalanche length, there was also a significant main effect of task condition (permutation ANOVA, $p < 0.05$). This capacity for consistent reorganization between successful and unsuccessful trials was previously reported by Corsi et al. (2024)[27], using a functional connectivity matrix derived from neuronal avalanches, known as the Avalanche Transition Matrix (ATM). We hypothesize that the brain network dynamics captured by the ATM could, in the future, serve as a novel source of features to support and extend our current findings—offering new, yet unexplored insights into brain activity during BCI tasks. These findings strongly support the conclusion that both neuronal avalanche length and activation are directly associated with distinct brain states—Rest and Motor Imagery—and are sensitive to the subject's ability to successfully perform the assigned BCI task. In other words, these features not only capture brain state differences but also reflect task performance at the trial level (see Supplementary Materials Figure 3). An additional and promising insight emerged from comparing ROI occurrence across all trials versus only Hit trials. This analysis, which underlies our ROI selection process, revealed a marked increase in T-values activations (Rest vs MI) within motor-related ROIs using only Hit trials. This supports the assumption that increased occurrence in motor areas is linked to successful motor imagery performance, thereby validating the rationale behind our ROI selection strategy (not show here).

*4.2  Repeated Measures Correlation and BCI-score prediction*

Building on this foundation, we assessed the relationship between specific neural features and BCI performance across sessions. Using repeated-measures correlation, we found that changes in avalanche length and activation—specifically the difference between motor imagery (MI) and resting-state conditions (Δ = MI – Rest)—are significantly correlated with individual BCI scores. While features during the Rest condition remained stable across sessions, those in the MI condition increased progressively, resulting in a shift from Δ values being negative (Rest > MI) to positive (MI > Rest). This shift was positively associated with improved BCI control. These findings were further supported by probability density function (PDF) analyses, which confirmed that greater MI feature values correspond to better BCI scores, both with and without ROI selection.

Despite advances in this area, many existing predictive models rely on single-session data, which fails to capture the evolving dynamics of learning. Some approaches have included stimulation-based predictors, such as transcranial magnetic stimulation (Pichiorri et al., 2011)[44] and median nerve stimulation (Marissens-Cueva, 2025)[45], but these too are often limited to short-term or one-off assessments. A notable exception is the work by Ma et al. (2022)[46], who used a multi-day EEG dataset to predict MI-BCI performance using both MI and resting-state data. However, their models were trained exclusively on first-session data, limiting their ability to account for longitudinal learning effects.

In contrast, our study demonstrates that robust and reliable prediction of BCI performance is achievable using data from as few as three sessions. By incorporating both resting-state and task-related neural activity, and applying repeated-measures correlation, we can predict fourth-session performance with high accuracy. This approach provides a more dynamic and longitudinal understanding of user learning, representing a significant step forward in adaptive BCI modelling.

Given the presence of significant repeated correlations using the same parameter combinations that also showed robust task condition and learning effects, we took an additional step beyond correlation analysis: we employed our features to predict BCI scores for each individual subject. From a clinical perspective, the ability to forecast BCI performance one session in advance would be highly valuable. This predictive capability is possible because our features encode both brain state discrimination (Rest vs. MI) and learning-related changes over time.
To simultaneously account for these two aspects, we implemented longitudinal models—specifically, Longitudinal Support Vector Regression (LSVR) and Longitudinal Support Vector Classification (LSVC). These models significantly outperformed their non-longitudinal counterparts (SVR and SVC), demonstrating the advantage of incorporating temporal structure into the analysis. Furthermore, results from session-shuffling experiments confirmed the importance of session order, reinforcing the relevance of modelling longitudinal progression.
The consistency of results before and after ROI selection supports our earlier hypothesis: it is feasible to reduce the dataset dimensionality via targeted ROI selection without compromising model performance. In fact, this reduction removes background noise and yields a computational benefit, which is particularly important in real-time BCI applications.

*4.3 Generalized Parameters Optimization*
Although differences and specific analytical strategies led to some variability in optimal parameter combinations, our analyses identified a subset of parameter settings that consistently enabled robust detection of neuronal avalanches across subjects, features, and sessions. Among the various z-threshold and minimum avalanche duration pairings tested—each grounded in neurophysiological plausibility—we found that several combinations effectively captured avalanche features (i.e., avalanche length and activations count) that were not only sensitive to task-related brain dynamics but also significantly correlated with and predictive of BCI performance. Despite each feature exhibiting distinct optimal parameter settings, we observed converging evidence supporting the use of z-thresholds set at the mean plus 2 to 3 standard deviations and minimum avalanche durations of 5 or 50 ms. These settings strike a balance between physiological interpretability and predictive utility, yielding avalanche dynamics that reflect meaningful brain state changes over time. While we acknowledge the nuanced distinctions between feature-specific optima, we propose this parameter range as a general and reliable guideline for applying neuronal avalanche analysis within the BCI context.

### 4.4  Limitations and Future Directions

To the best of our knowledge, this study is the first to apply longitudinal models in the context of BCI training with the goal of predicting individual BCI performance one session in advance. However, a major limitation lies in the relatively small number of training sessions, which is a linear learning trajectory across sessions, which may not accurately reflect the complex and variable nature of learning processes in real-world settings. Factors such as physiological fluctuations and environmental influences—both of which can affect daily performance—are not explicitly accounted for in the current model.

Nonetheless, our findings provide a promising framework that could support clinicians in forecasting individual progress and designing personalized training protocols. For some users, fewer sessions may suffice to gain BCI control, while others may require extended training. Tailoring the number and structure of training sessions to each user could help reduce frustration and improve the likelihood of successful BCI adoption. In future work, we aim to validate our approach using datasets that include a larger number of training sessions.

This would allow us not only to improve the prediction of BCI performance one or more sessions ahead, but also to estimate the total number of sessions each subject might need to achieve effective control. We also plan to explore different classification thresholds to refine the prediction process. Moreover, we are currently working on enhancing the robustness of the ROI selection process across various parameter combinations. Our goal is to define a stable, group-level set of ROIs that remains consistent across subjects while preserving the core principle of identifying regions that encode both the distinction between Rest and MI brain states and their evolution over different training sessions. Achieving a unified and interpretable ROI framework will further strengthen the clinical applicability and scientific insight of our findings after the ROIs selection dataset reduction. To achieve this goal, we think that studying the functional connectivity structured like an Avalanche Transition Matrix could capture deeper mechanism and brain activity behaviour underlined the neuronal avalanches thanks to specific propriety of the generated brain network.

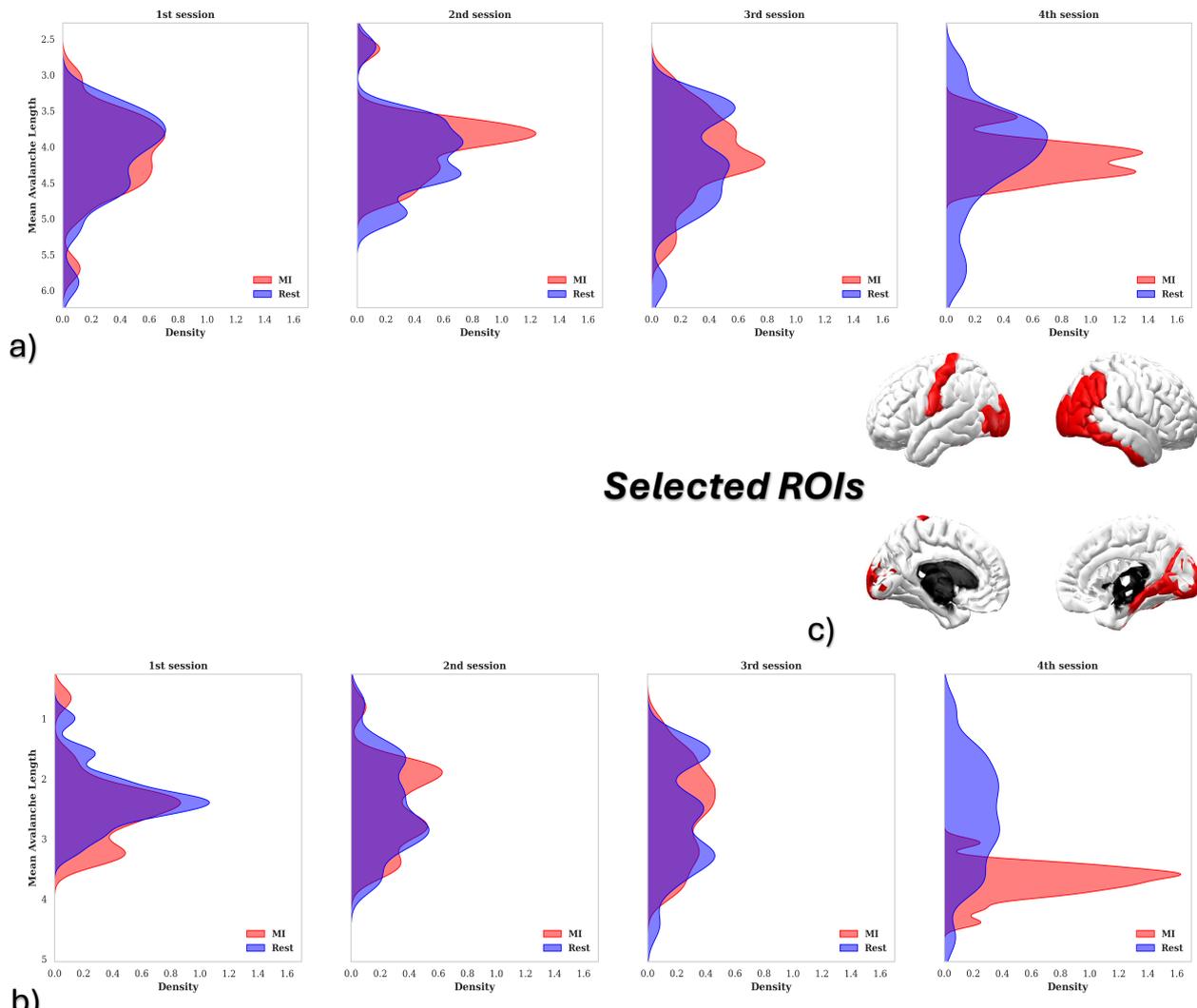

*Figure 9: Probability Density function (PDF) : PDF of mean avalanche length across all the subjects, repeated for all the training sessions (from the first session on the left to the fourth session on the right) to rappresent the different distribution in MI task condition (red) and in resting state (blue). It's computed over the entire dataset **(a)** and over a selected set of ROIs **(c)** in **(b)**.*

## *Conclusion*

Our findings demonstrate that neuronal avalanches are highly effective in distinguishing between resting-state and motor imagery brain activity. Moreover, they can reliably indicate whether a subject successfully performs the assigned task. These results highlight, also, the potential of neuronal avalanches as powerful biomarkers for tracking and supporting the BCI training process. In this study, we introduce two novel features—avalanche length and activation—that offer a richer and more comprehensive representation of brain dynamics compared to traditional BCI features. While each feature demonstrates slightly different sensitivities to specific parameter combinations, avalanche length consistently outperformed in detecting task-related effects, capturing learning dynamics, correlating with BCI performance, and achieving higher prediction accuracy.

Finally, this work proposes a promising strategy to address the issue of BCI inefficiency by supporting the development of personalized training programs. Tailoring the training process to individual learning profiles may help clinicians enhance user engagement, reduce frustration, and ultimately increase the number of successful BCI users.

*References :*

1.  Thompson, M. C. Critiquing the Concept of BCI Illiteracy. *Sci. Eng. Ethics* **25**, 1217–1233 (2019).

2.  Dadarlat, M. C., Canfield, R. A. & Orsborn, A. L. Neural Plasticity in Sensorimotor Brain–Machine Interfaces. *Annu. Rev. Biomed. Eng.* **25**, 51–76 (2023).

3.  Alkoby, O., Abu-Rmileh, A., Shriki, O. & Todder, D. Can We Predict Who Will Respond to Neurofeedback? A Review of the Inefficacy Problem and Existing Predictors for Successful EEG Neurofeedback Learning. *Neuroscience* **378**, 155–164 (2018).

4.  Nijboer, F. *et al.* A P300-based brain–computer interface for people with amyotrophic lateral sclerosis. *Clin. Neurophysiol.* **119**, 1909–1916 (2008).

5.  Nijboer, F., Birbaumer, N. & Kubler, A. The Influence of Psychological State and Motivation on Brain–Computer Interface Performance in Patients with Amyotrophic Lateral Sclerosis – a Longitudinal Study. *Front. Neurosci.* **4**, (2010).

6.  Guger, C., Edlinger, G., Harkam, W., Niedermayer, I. & Pfurtscheller, G. How many people are able to operate an EEG-based brain-computer interface (BCI)? *IEEE Trans. Neural Syst. Rehabil. Eng.* **11**, 145–147 (2003).

7.  Witte, M., Kober, S. E., Ninaus, M., Neuper, C. & Wood, G. Control beliefs can predict the ability to up-regulate sensorimotor rhythm during neurofeedback training. *Front. Hum. Neurosci.* **7**, (2013).

8.  Hardman, E. *et al.* Frontal interhemispheric asymmetry: self regulation and individual differences in humans. *Neurosci. Lett.* **221**, 117–120 (1997).

9.  Hammer, E. M. *et al.* Psychological predictors of SMR-BCI performance. *Biol. Psychol.* **89**, 80–86 (2012).

10. Vuckovic, A. & Osuagwu, B. A. Using a motor imagery questionnaire to estimate the performance of a Brain–Computer Interface based on object oriented motor imagery. *Clin. Neurophysiol.* **124**, 1586–1595 (2013).

11. Blankertz, B. *et al.* Neurophysiological predictor of SMR-based BCI performance. *NeuroImage* **51**, 1303–1309 (2010).

12. Halder, S. *et al.* Neural mechanisms of brain–computer interface control. *NeuroImage* **55**, 1779–1790 (2011).

13. Guillot, A. *et al.* Functional neuroanatomical networks associated with expertise in motor imagery. *NeuroImage* **41**, 1471–1483 (2008).

14. Halder, S. *et al.* Prediction of Auditory and Visual P300 Brain-Computer Interface Aptitude. *PLoS ONE* **8**, e53513 (2013).

15. Enriquez-Geppert, S. *et al.* Modulation of frontal-midline theta by neurofeedback. *Biol. Psychol.* **95**, 59–69 (2014).

16. Sannelli, C., Vidaurre, C., Müller, K.-R. & Blankertz, B. A large scale screening study with a SMR-based BCI: Categorization of BCI users and differences in their SMR activity. *PLoS ONE* **14**, e0207351 (2019).

17. Grosse-Wentrup, M., Mattia, D. & Oweiss, K. Using brain–computer interfaces to induce neural plasticity and restore function. *J. Neural Eng.* **8**, 025004 (2011).

18. Maeder, C. L., Sannelli, C., Haufe, S. & Blankertz, B. Pre-Stimulus Sensorimotor Rhythms Influence Brain–Computer Interface Classification Performance. *IEEE Trans. Neural Syst. Rehabil. Eng.* **20**, 653–662 (2012).

19. Weber, E., Köberl, A., Frank, S. & Doppelmayr, M. Predicting Successful Learning of SMR Neurofeedback in Healthy Participants: Methodological Considerations. *Appl. Psychophysiol. Biofeedback* **36**, 37–45 (2011).

20. Mohanty, R., Sethares, W. A., Nair, V. A. & Prabhakaran, V. Rethinking Measures of Functional Connectivity via Feature Extraction. *Sci. Rep.* **10**, 1298 (2020).

21. Gonzalez-Astudillo, J., Cattai, T., Bassignana, G., Corsi, M.-C. & Fallani, F. D. V. Network-based brain–computer interfaces: principles and applications. *J. Neural Eng.* **18**, 011001 (2021).

22. Brake, N. *et al.* A neurophysiological basis for aperiodic EEG and the background spectral trend. *Nat. Commun.* **15**, 1514 (2024).

23. Beggs, J. M. & Plenz, D. Neuronal Avalanches in Neocortical Circuits. *J. Neurosci.* **23**, 11167–11177 (2003).

24. Arviv, O., Goldstein, A. & Shriki, O. Neuronal avalanches and time-frequency representations in stimulus-evoked activity. *Sci. Rep.* **9**, 13319 (2019).

25. Sorrentino, P. *et al.* The structural connectome constrains fast brain dynamics. *eLife* **10**, e67400 (2021).

26. Rabuffo, G., Fousek, J., Bernard, C. & Jirsa, V. Neuronal Cascades Shape Whole-Brain Functional Dynamics at Rest. *eNeuro* **8**, ENEURO.0283-21.2021 (2021).

27. Corsi, M.-C. *et al.* Measuring neuronal avalanches to inform brain-computer interfaces. *iScience* **27**, 108734 (2024).

28. Mannino, Camilla, Sorrentino, Pierpaolo, Chavez, Mario, & Corsi, Marie-Constance. NEURONAL AVALANCHES FOR EEG-BASED MOTOR IMAGERY BCI. doi:10.3217/978-3-99161-014-4-018.

29. Corsi, M.-C. *et al.* Functional disconnection of associative cortical areas predicts performance during BCI training. *NeuroImage* **209**, 116500 (2020).

30. Sorrentino, P. *et al.* The structural connectome constrains fast brain dynamics. *eLife* **10**, e67400 (2021).

31. Hannah, R., Cavanagh, S. E., Tremblay, S., Simeoni, S. & Rothwell, J. C. Selective Suppression of Local Interneuron Circuits in Human Motor Cortex Contributes to Movement Preparation. *J. Neurosci.* **38**, 1264–1276 (2018).

32. Endo, H., Kizuka, T., Masuda, T. & Takeda, T. Automatic activation in the human primary motor cortex synchronized with movement preparation. *Cogn. Brain Res.* **8**, 229–239 (1999).

33. Bakdash, J. Z. & Marusich, L. R. Repeated Measures Correlation. *Front. Psychol.* **8**, (2017).

34. Müller-Putz, G. R., Scherer, R., Brunner, C., Leeb, R. & Pfurtscheller, G. Better than random? A closer look on BCI results.

35. Awad, M. & Khanna, R. Support Vector Regression. in *Efficient Learning Machines: Theories, Concepts, and Applications for Engineers and System Designers* (eds. Awad, M. & Khanna, R.) 67–80 (Apress, Berkeley, CA, 2015). doi:10.1007/978-1-4302-5990-9_4.

36. Mammone, A., Turchi, M. & Cristianini, N. Support vector machines. *WIREs Comput. Stat.* **1**, 283–289 (2009).

37. Du, W. *et al.* A longitudinal support vector regression for prediction of ALS score. in *2015 IEEE International Conference on Bioinformatics and Biomedicine (BIBM)* 1586–1590 (2015). doi:10.1109/BIBM.2015.7359912.

38. Chen, S. & Bowman, F. D. A Novel Support Vector Classifier for Longitudinal High-dimensional Data and Its Application to Neuroimaging Data. *Stat. Anal. Data Min.* **4**, 604–611 (2011).

39. Ahn, M. & Jun, S. C. Performance variation in motor imagery brain–computer interface: A brief review. *J. Neurosci. Methods* **243**, 103–110 (2015).

40. Ahn, M., Cho, H., Ahn, S. & Jun, S. C. High Theta and Low Alpha Powers May Be Indicative of BCI-Illiteracy in Motor Imagery. *PLoS ONE* **8**, e80886 (2013).

41. Stiso, J. *et al.* Learning in brain-computer interface control evidenced by joint decomposition of brain and behavior. *J. Neural Eng.* **17**, 046018 (2020).

42. Jeunet, C., Jahanpour, E. & Lotte, F. Why standard brain-computer interface (BCI) training protocols should be changed: an experimental study. *J. Neural Eng.* **13**, 036024 (2016).

43. Vidaurre, C., Krämer, N., Blankertz, B. & Schlögl, A. Time Domain Parameters as a feature for EEG-based Brain–Computer Interfaces. *Neural Netw.* **22**, 1313–1319 (2009).

44. Pichiorri, F. *et al.* Brain–computer interface boosts motor imagery practice during stroke recovery. *Ann. Neurol.* **77**, 851–865 (2015).

45. Marissens Cueva, V., Bougrain, L., Lotte, F. & Rimbert, S. Reliable predictor of BCI motor imagery performance using median nerve stimulation. *J. Neural Eng.* **22**, 026039 (2025).

46. A large EEG dataset for studying cross-session variability in motor imagery brain-computer interface | Scientific Data. https://www.nature.com/articles/s41597-022-01647-1.

*Supplementary Materials:*

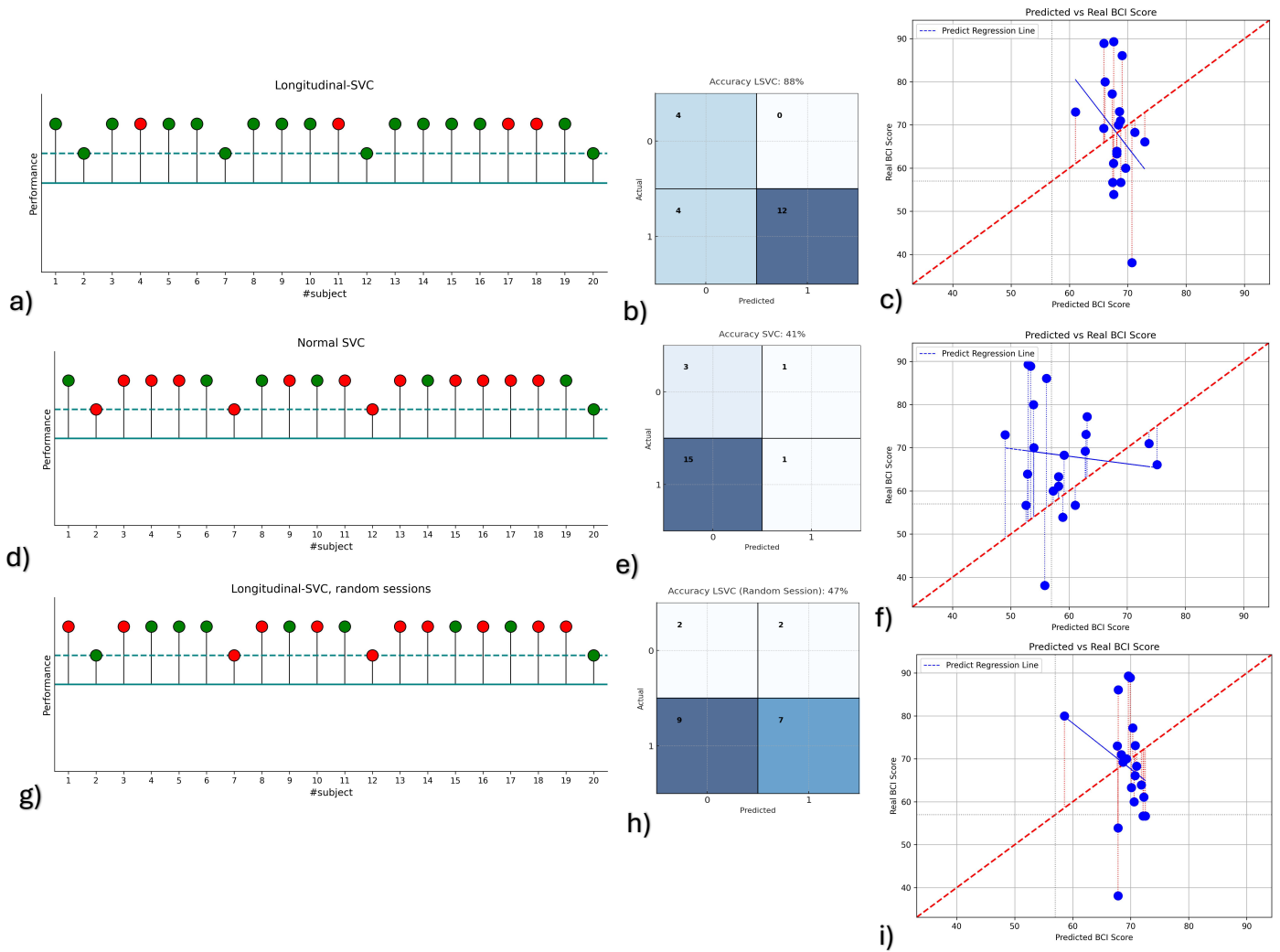

*Supplementary Materials, Figure 1 : Predictive model results*
*(a, d, g)* **Classification results:** *Each ball represents one subject. The dashed line indicates the classification threshold; subjects above the line are considered able to control the outcome. Green balls indicate correct predictions of control ability, while red balls represent misclassifications. Panel (a) shows results using a longitudinal predictive model, which are compared to a standard SVC model in panel (d) and to use LSVC with a random shuffle of the training sessions (g).*
*(b,e,h)* **Confusion Matrix of classification results:** *Confusion Matrix associated to LSVC (b), Confusion Matrix associated to SVC (e) and Confusion Matrix associated to LSVC with a random shuffle of the training sessions (h)*
*(c,f,i)* **Regression results:** *Each point represents one subject. The red dashed lines indicate the prediction error between actual and predicted values. The bold red line shows the optimal prediction trend, while blue lines represent the predicted trends. Panel (c) shows results from a longitudinal regression model, compared to a standard SVR model in panel (f) and to use LSVR with a random shuffle of the training sessions.*
***All these plots are realized using z-threshold : µ + 3σ, min. avalanches length: 12, best coupled parameters for prediction***

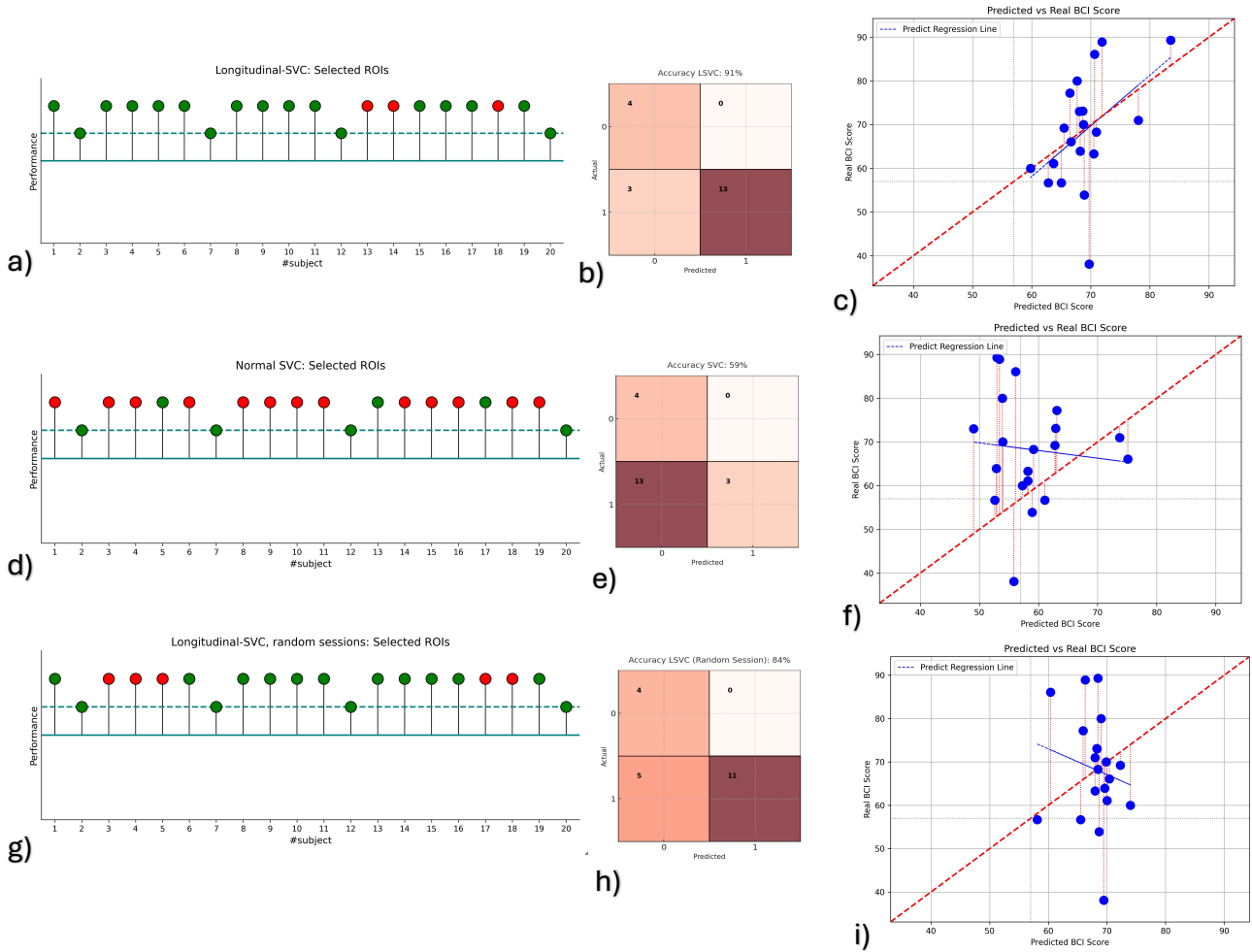

***Supplementary Materials, Figure 2 : Predictive model results on a selected set of ROIs***
***(a, d, g) Classification results:*** *Each ball represents one subject. The dashed line indicates the classification threshold; subjects above the line are considered able to control the outcome. Green balls indicate correct predictions of control ability, while red balls represent misclassifications. Panel (a) shows results using a longitudinal predictive model, which are compared to a standard SVC model in panel (d) and to use LSVC with a random shuffle of the training sessions (g).*
***(b,e,h) Confusion Matrix of classification results:*** *Confusion Matrix associated to LSVC (b), Confusion Matrix associated to SVC (e) and Confusion Matrix associated to LSVC with a random shuffle of the training sessions (h)*
***(c,f,i) Regression results:*** *Each point represents one subject. The red dashed lines indicate the prediction error between actual and predicted values. The bold red line shows the optimal prediction trend, while blue lines represent the predicted trends. Panel (c) shows results from a longitudinal regression model, compared to a standard SVR model in panel (f) and to use LSVR with a random shuffle of the training sessions.*
***All these plots are realized using z-threshold : μ + 2σ, min. avalanches length: 12, best coupled parameters for prediction***

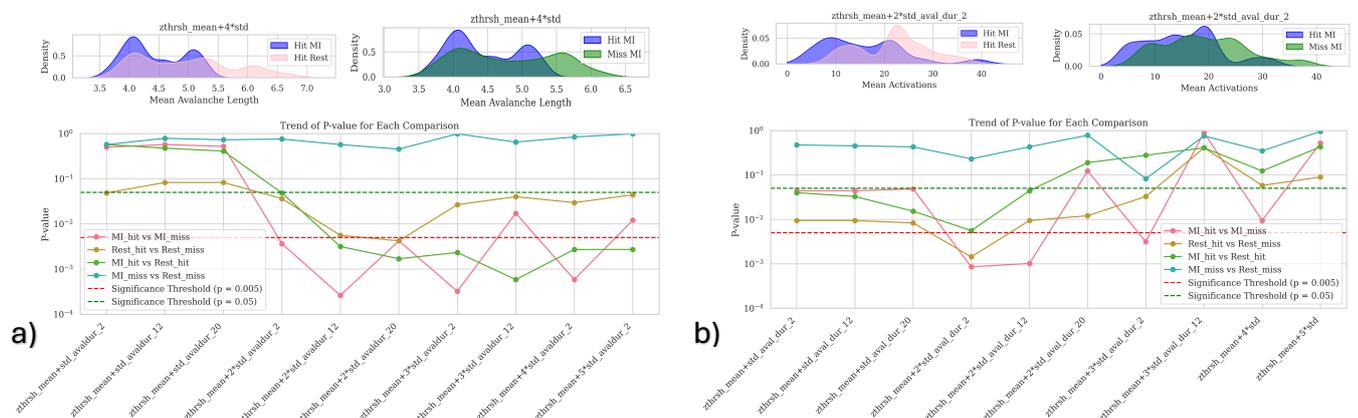

***Supplementary Materials, Figure 3: a) Mean avalanches length hit trials*** *(when subject is able to well perform the task)* ***vs. miss trial*** *(when subject is not able to perform the task): on the top the probability density function that represent Hit MI trials (in blue) vs. Hit Rest trials (in pink) [on the left] and Hit MI trials (in blue) vs. Miss MI trials (in green) [on the right]; bellow: Trends of p-values from pairwise comparisons of mean avalanche length across brain states [MI vs. Rest] and conditions [Hit vs. Miss] across possible coupled parameters. Dashed horizontal lines indicate significance thresholds (p < 0.05, Bonferroni-corrected p < 0.0011).* ***b) Activations hit trials*** *(when subject is able to well perform the task)* ***vs. miss trial*** *(when subject is not able to perform the task): on the top the probability density function that represent Hit MI trials (in blue) vs. Hit Rest trials (in pink) [on the left] and Hit MI trials (in blue) vs. Miss MI trials (in green) [on the right]; bellow: Trends of p-values from pairwise comparisons of activations count across brain states [MI vs. Rest] and conditions [Hit vs. Miss] across possible coupled parameters. Dashed horizontal lines indicate significance thresholds (p < 0.05, Bonferroni-corrected p < 0.0011).* ***All these analysis are perform only on the last training session.***